\documentclass[pdftex,twocolappendix,numberedappendix,appendixfloats,apj]{emulateapj}
\usepackage{color}
\usepackage{amsmath}

\begin{document}
\shorttitle{3D Radiation hydrodynamics of a dynamical torus}

\shortauthors{Williamson, Venanzi \& H\"{o}nig}

\title{3D Radiation hydrodynamics of a dynamical torus}

\author{David Williamson}
\affil{Department of Physics \& Astronomy, University of Southampton, Southampton, SO17 1BJ, UK}
\email{d.j.williamson@soton.ac.uk}

\author{Marta Venanzi}
\affil{Department of Physics \& Astronomy, University of Southampton, Southampton, SO17 1BJ, UK}
\email{marta.venanzi@gmail.com}

\and

\author{Sebastian H\"{o}nig}
\affil{Department of Physics \& Astronomy, University of Southampton, Southampton, SO17 1BJ, UK}
\email{s.hoenig@soton.ac.uk}

\begin{abstract}
We have developed a new dynamical model of the torus region in active galactic nucleus (AGN), using a three-dimensional radiation hydrodynamics algorithm. These new simulations have the specific aim to explore the role of radiatively-driven outflows, which is hotly debated in current literature as a possible explanation for the observed infrared emission from the polar regions of AGN. In this first paper, we only consider radiative effects induced by the primary radiation from the AGN. The simulations generate a disk \& outflow structure that qualitatively agrees with observations, although the outflow is radial rather than polar, likely due to the lack of radiation pressure from hot dust. We find cut-offs between the wind and disk at gas temperatures of $1000$ K and dust temperatures of $100$ K, producing kinematic signatures that can be used for interpretation of high resolution infrared observations. We also produce line emission maps to aid in the interpretation of recent ALMA observations and future JWST observations. We investigate a number of simulation parameters, and find that the anisotropy of the radiation field is equally important to the Eddington factor, despite the anisotropy often being assumed to have a single sometimes arbitrary form in many previous works. We also find that supernovae can have a small but significant impact, but only at extremely high star formation rates.
\end{abstract}


\section{Introduction}\label{section_intro}

What is the physical origin and the dynamical state of the ``torus'' in active galactic nuclei (AGN)? While the term implies a specific geometry, it should be merely considered a place holder for the optically thick region that surrounds the bright core of an active galactic nucleus (AGN) in the standard unification model \citep{1993ARA&A..31..473A}. Dust in this region blocks the radiation from the bright central accretion disk and broad line region (BLR) when viewed edge-on, explaining the differences between type 1 and type 2 AGN as orientation effects. Observations have further shown that this obscuring material consists of potentially clumpy, dusty gas, as evidenced by the emission of infrared (IR) radiation roughly in thermal equilibrium with the AGN \citep[for recent reviews see][]{2013arXiv1301.1349H,2015ARA&A..53..365N}. However, determining the origin and physical state, and how such material can cover such a large solid angle has proved challenging.

The thickness of the torus can not be explained through thermal pressure, as the required high temperatures ($T>10^6$ K) would destroy any dust \citep{1988ApJ...329..702K}. While a dynamical system of dust clouds with a large velocity dispersion could provide the required covering fraction \citep{1992ApJ...401...99P,1995MNRAS.272..737R,2004A&A...426..445B}, it is not clear if such a structure would be dynamically and physically stable, as the cloud collision rate would be large. The clouds would convert their kinetic energy into thermal energy and cool into a flat disk \citep{1988ApJ...329..702K}, with the clouds potentially disintegrating and/or dissipating. Warped disks, misaligned disks, and disordered flows of dusty gas have also been proposed as a possible solutions, but the model disks are either extremely large \citep{1989ASIC..290..457P,1989ApJ...347...29S}, or have an insufficient covering fraction \citep{2012MNRAS.420..320H}. 

\citet{2007ApJ...661...52K} argued that the vertical height may be maintained by IR radiation pressure from AGN photons that have been reprocessed by optically thick dust. This analytic model was revisited in \citet{2016ApJ...825...67C} and \citet{2017ApJ...843...58C} in 3D radiation hydrodynamic (RHD) and radiation magnetohydrodynamic (RMHD) simulations using the reduced speed-of-light approximation to solve the radiation equations. In these later simulations, a more dynamical picture emerged. At modestly low Eddington ratios representative of Seyfert galaxies, IR radiation pressure was found to be insufficient to fully support the thickness of the dusty gas at all radii. A strong radial wind was produced, peeling off the inner layer of the torus and pushing the gas outwards. In these simulations, the AGN emission was assumed to be isotropic, which may over-estimate the radiation support of the torus by providing a relatively large number of photons in the plane as compared to the polloidally concentrated radiation field expected from an accretion disk.

The long-standing problem of the torus vertical scale height has acquired renewed attention when IR interferometry found that many nearby AGN show parsec-scale dust emission originating from the polar region of the AGN, rather than the equatorial region classically identified with the torus \citep{2012ApJ...755..149H,2013ApJ...771...87H,2014A&A...563A..82T,2016A&A...591A..47L,2018ApJ...862...17L}. With a geometrically thick torus already being difficult to explain in the current theoretical framework, the elevation of dust to large heights above the disk \citep[possibly up to 100\,pc scales, e.g.][]{2016ApJ...822..109A} poses an even bigger challenge. Phenomenological radiative transfer models suggest that both the distribution of the resolved IR emission and the broad IR SED are best reproduced with two components: a geometrically thin, dense dusty disk reaching from the sublimation radius outwards and a hollow cone originating near the inner edge of the dusty disk \citep{2017ApJ...838L..20H,2017MNRAS.472.3854S}. 

Hence, RHD models began focusing on the role of radiative processes as a way to drive dusty winds. \citet{2015ApJ...812...82W} presented a radiation-driven fountain model on a scale larger than the resolved IR emission (a $32$ pc box). In this model, the dust distribution is not static, but rather a dynamic ``fountain'' launched from a dusty disk. The fountain either falls back onto the disk or is dissipated into the interstellar medium at distances of several parsecs. As these simulations emphasize the role of a wind driven by anisotropic UV radiation pressure and X-ray heating, they neglect IR radiation pressure.
The disk is inflated by gravitational instabilities driven by explicitly modeled self-gravity and feedback from star-formation through supernovae \citep{2016ApJ...828L..19W}. However, the cost of such a large domain is a coarse resolution of $0.125$ pc.

\citet{2016MNRAS.460..980N} performed 2-dimensional RHD simulations at high resolution, focusing on the inner $1.2$ pc), including the effects of metal cooling, self-gravity, anisotropic X-ray photons, and multigroup IR radiation transfer in the single-scattering domain. They found that the metal cooling effectively suppressed the vertical support from IR radiation pressure against collapse. However, a wind forms that could give the expected covering fractions, leading to a disk and wind configuration, matching the proposed explanation of results from IR interferometry \citep{2012ApJ...755..149H}. They found that the angle of the wind depends on the dust grain size and the presence of scattering. These simulations were performed with a very high Eddington factor of $0.77$, beyond the regime of the majority of nearby AGN that showed polar elongation in IR interferometry.

\citet{2016ApJ...819..115D} presented a 2-dimensional torus model with isotropic radiation, where high energy radiation is ray-traced while IR radiation is treated with flux-limited diffusion (FLD). This model produces a strong wind, but retains a thick torus, potentially due to the use of the FLD method, which effectively acts as a non-thermal source of pressure, puffing up the torus and smoothing out structure. 

All these simulations use different modelling strategies, and often come to drastically different conclusions. This leaves the nature of the torus and wind essentially unconstrained, as the results of simulations depend not only on physical parameters (e.g. the dust population, supernovae, etc.) and the SED and anisotropy of the primary AGN radiation, but also on the numerical implementation (or inclusion) of radiative transfer, metal cooling, and self-gravity, as well as basic numerical issues such as resolution.

To determine how to proceed, we turn to the recent observations which lend support to a dynamical model where many of the observed torus properties, including angle-dependent obscuration, are the result of a dusty wind or outflow. In this work, we introduce a novel 3D RHD model that approaches the numerical issues from a unique angle, while focusing on the wind as the source of dusty opacity and emission. The basic picture is a cool dense disk around the BLR, which develops a wind due to a combination of the heating and radiation pressure from the AGN, from the stellar field, and from its own IR emission. As we are primarily concerned with the generation of the wind and its properties, we only model the central part of the cool disk to ensure maximum resolution. The goal of this model is to produce a disk wind that contains hot dust and produces the polar extended mid-IR emission.

In this paper, we only consider radiation from the AGN and the diffuse stellar field, leaving the more computationally complex problem of scattering and IR re-radiation to our next paper. We use Cloudy \citep{2013RMxAA..49..137F,2017RMxAA..53..385F} to incorporate heating, cooling, and radiation pressure on gas and a physically consistent composition of dust, including differential sublimation. The most unique element of our approach is the use of a Lagrangian formulation for hydrodynamics, directly providing very high resolution in dense regions while simultaneously capturing large length scales accurately. This allows us to efficiently resolve the (column) density regime where winds are launched.

In Section~\ref{section_method} we describe our method, its advantages and approximations, and our suite of simulations. In Section~\ref{section_results} we describe the evolution of the simulations, and their observable properties. In Section~\ref{section_discussion} we discuss our results in context of the literature, and we summarize our conclusions and briefly discuss their implications in Section~\ref{section_conclusion}

\section{Method}\label{section_method}

\subsection{Simulation Code}

We use the public version of the N-body+hydrodynamics code GIZMO \citep{2015MNRAS.450...53H} in pressure smoothed-particle hydrodynamics (P-SPH) mode. GIZMO's P-SPH algorithm is a `modern' SPH method as opposed to the `traditional' T-SPH mode. P-SPH is a pressure-entropy scheme, incorporating artificial conductivity, and switches for artificial viscosity. GIZMO is a descendant of GADGET \citep[specifically GADGET-3, see][]{2005MNRAS.364.1105S}, an N-body+SPH code in widespread use, and GIZMO shares many of the same features with GADGET, including file formats. GIZMO primarily differs from GADGET by including a new hydrodynamics scheme that differs significantly from SPH, but in our initial tests we found this new algorithm sometimes developed extremely small time-steps that essentially halted the simulation. It was not clear to us whether the cause of this issue was a bug in the particular public version of GIZMO we download (a bug that may have been fixed in a more recent release), a genuine issue with the algorithm, or a problem in our particular model and its implementation, but recent results show that many of the results of cosmological simulations do not strongly depend on the hydrodynamics scheme \citep{2018MNRAS.480..800H}. Hence rather than attempting to disentangle this problem, we chose to run GIZMO in the simpler P-SPH mode, where we did not encounter such vanishingly small time-steps.

Our simulations include self-gravity, which can produce gravitational instabilities, as we discuss in Section~\ref{section_evolve}. Additionally, we include a static background potential, which we discuss in Section~\ref{simulations_section}.

We summarize the key points of our method in the subsections below.

\subsection{Hydrodynamics}

The previous simulations of the AGN torus region mentioned in Section~\ref{section_intro} were performed with an Eulerian formulation on a fixed grid in two or three dimensions. We differ from the standard approach by using SPH, a Lagrangian formulation of hydrodynamics. Eulerian methods are limited by length resolution, and tend to resolve high density fluid poorly, while Lagrangian methods are limited by mass resolution, and tend to resolve low density fluid poorly. Eulerian methods suffer from an inherent numerical diffusivity (in addition to an explicit numerical viscosity), which causes fluid to spread out, especially at high velocities or when the velocity field is not aligned with the grid, producing errors in high-velocity outflows. On the other hand, SPH methods include an explicit numerical viscosity term to capture shocks, which dissipates energy and causes fluid to collapse, as well as suppressing fluid instabilities. There are various techniques to alleviate the problems with either method, such as adaptive-mesh refinement and particle splitting, but these remain only partial solutions. Essentially, SPH and grid methods have complementary strengths and weaknesses, and by choosing to use a Lagrangian method, we are exploring the problem from an almost orthogonal perspective to existing studies. In this particular context, our model will not effectively resolve low-density winds in the ionisation cone of the AGN. However, it will efficiently resolve the denser gas that contributes the most to the opacity and infrared luminosity of the torus region.

Traditional SPH methods have a number of well-known problems with sharp density gradients, producing a spurious surface-tension effect, and suppressing turbulence and the Kelvin-Helmholtz and Rayleigh-Taylor instabilities \citep{2007MNRAS.380..963A}. Several modifications to traditional SPH have been proposed to alleviate these effects \citep[see][and references therein]{2013MNRAS.428.2840H}. The pressure-entropy formulation \citep{2013MNRAS.428.2840H,2013ApJ...768...44S} explicitly calculates pressure from the smoothing kernel rather than from particle densities. This removes the spurious forces that result from sharp density contrasts, provided that the physical pressure gradient is smooth. An artificial viscosity is still required to capture shocks, where the pressure gradient is very steep. This method is inherently adiabatic in the sense that entropy is manifestly conserved in the equations of motion, while energy conservation is only subject to very small errors. Note that, with the appropriate choice of definitions, the pressure-entropy formulation is equivalent to a `pressure-energy' formulation.

The basic equations of motion to be solved are the compressible Euler equations. Mass is conserved trivially in SPH, and so we need only solve the momentum and energy conservation equations:
\begin{align}
\frac{D\mathbf{v}}{Dt}&=-\frac{\nabla P}{\rho} + \mathbf{g} + \mathbf{a_r},\\
\frac{Du}{Dt}&=-\frac{P}{\rho}\nabla\cdot\mathbf{v} + H,\\
\end{align}
where $D/Dt$ is the convective derivative, $P$, $\rho$, $u$, and $\mathbf{v}$ are the pressure, density, internal energy, and velocity of the fluid, $\mathbf{g}$ is the acceleration from gravity, $\mathbf{a_r}$ is the radiative acceleration, and $H$ is the combination of all heating and cooling effects. The base equations without the source terms ($\mathbf{g}$,$\mathbf{a_r}$,$H$) are discretized in the code following the pressure-entropy formulation in \citet{2013MNRAS.428.2840H}. The source terms are then added to the purely hydrodynamic rates, which are integrated with a leapfrog scheme. Heating and cooling rates can vary rapidly with temperature, and so to avoid short system time-steps we integrate heating and cooling over shorter sub-steps, and apply the time-averaged rate to $Du/Dt$.

We assume an ideal equation of state, i.e.
\begin{equation}
P=(\gamma-1)u\rho
\end{equation}where the adiabatic constant $\gamma=5/3$. Temperature is only calculated for Cloudy table lookup and for visualisation, and is calculated through
\begin{equation}
T=(\gamma-1)\frac{\mu m_p}{k_B}u,
\end{equation}
where $k_B$ is the Boltzmann constant, $m_p$ is the proton mass, and $\mu\approx0.122$ is the mean molecular mass, i.e. $\rho/n_\mathrm{H}$.

\subsection{Radiative Transfer}\label{section_rad}

We have implemented a raytracing algorithm within GIZMO that calculates the optical depth between the central AGN and each gas particle. Raytracing, even from a single point, can be computationally expensive -- a naive algorithm scales with the square of the number of particles, $N^2$, as there are $N$ particles receiving radiation, and $N$ particles to potentially contribute extinction. Our algorithm uses an oct-tree structure to efficiently identify which particles intersect each ray, and uses this to calculate the optical depth between the particle and the AGN. We use the algorithm of \citet{Revelles00anefficient} for efficient oct-tree traversal.

We have performed a suite of Cloudy models \citep{2013RMxAA..49..137F,2017RMxAA..53..385F} to tabulate the internal optical depth, heating, cooling, and radiation pressure of dusty gas as a function of the gas temperature, gas density, the optical depth from the gas particle to the AGN, and the intensity of incident AGN radiation. The Cloudy models also include a radiation field from bulge stars (see Section~\ref{appendix_cloudy}). For each particle, we identify the particles that intersect a ray from the AGN to the particle's centre. For each intersecting particle, we calculate its contribution to the optical depth from its tabulated opacity and the intersecting column density, determined from the impact parameter through a cubic spline kernel of variable smoothing length.

That is, most of the details of radiative transfer are pre-calculated with Cloudy, and only the optical depths from the AGN to each particle are calculated in the code. Essentially, the radiative transfer equation is reduced to simply:
\begin{equation}
\zeta_i = \zeta_\mathrm{tab}(T_i,\rho_i,F_i,\tau_i),
\end{equation}
where $\zeta_i$ represents the radiative acceleration, heating/cooling rate, or opacity of particle $i$, and $\zeta_\mathrm{tab}(T_i,\rho_i,F_i,\tau_i)$ is that value interpolated from the pre-calculated Cloudy tables as a function of $T_i$, the particle temperature, $\rho_i$, the particle density, $F_i$, the unextinguished AGN flux, and $\tau_i$, the optical depth between the AGN and the particle. $F_i$ is defined below in section~\ref{anisotopic_flux}. $\tau_i$ is calculated as 
\begin{equation}
\tau_i = \sum_j \kappa_j \Sigma_j(\mathbf{r}_i,\mathbf{r}_\mathrm{AGN}),
\end{equation}
where the sum is over all particles that are close enough to perhaps intersect the ray between the AGN and particle $i$ (as determined by the oct-tree algorithm), $\kappa_j$ is the opacity (in area/mass units) of particle $j$ as determined from the table in the previous step, and $\Sigma_j(\mathbf{r}_i,\mathbf{r}_\mathrm{AGN})$ is the column density contributed by particle $j$ to a ray from the AGN to the centre of particle $i$, as calculated by the raytracing method.

\subsubsection{Resolution}

Most of the momentum deposition from the radiation pressure of the AGN is applied to the optically thin ($\tau\lesssim1$) ``skin'' of the gas. If this is not resolved -- that is, if the gas particles that receive AGN radiation are optically thick -- then the wind will be suppressed. Instead of force being applied to an optically thin skin which then peels away as a wind, this momentum would be distributed over the optical depth of the particle, diluting the acceleration of the gas. Furthermore, an optically thick particle can no longer be reasonably assumed to have a single value of temperature, opacity, or heating and cooling rates.

In the Lagrangian formulation, the central column density of an SPH particle is determined by its mass and smoothing length, where the smoothing length $h$ is adaptive to ensure each particle has enough neighbors. For a particle of fixed mass $m_p$, the column density is proportional to $m_{p}h^{-2}$. Hence at any mass resolution, for a small enough $h$, the column density for a particle is high enough that the particle is optically thick. The adaptive smoothing length is effectively a density criterion, which roughly ensures that $h\propto \rho^{-1/3}/m_{p}$. If we define the threshold for an optically thick column density to be $\Sigma_T$, the cutoff density is proportional to $\Sigma_T^{3/2}m_p^{-1/2}$, beyond which particles become optically thick.

Hence, while out-of-the-box SPH with adaptive smoothing and softening lengths can capture the dynamics of dense gas, such as that produced by gravitational collapse (although this becomes computationally expensive as densities become large), radiative processes will not be correctly treated. Additionally, Cloudy does not always produce converged results at high densities. 

We have taken several steps to alleviate these problems.
\begin{itemize}
  \item We only calculate Cloudy tables up to $n=10^9$~cm$^{-3}$, and extrapolate for higher $n$ values, by assuming the heating rate remains constant, but that the cooling rate per unit mass is proportional to density (that is, the volumetric cooling rate is proportional to $n^2$). This dense gas mostly occurs in dense cores, and the precise form of the extrapolation does not affect the nature of the wind, which is the focus of this study. To avoid over-accelerating distant gas, we also extrapolate the radiation pressure for fluxes below $F=10^{1.25}$ erg~s$^{-1}$~cm$^{-2}$ by assuming the radiation pressure is proportional to the flux.
  \item We set the star formation threshold to $n=10^{10}$ cm$^{-3}$, providing a sink for the very high density gas. Our star formation algorithm is explained more fully in Section~\ref{section_starform}, 
  \item We generate a new set of tables which account for self-shielding. The optical depth of a particle depends on its opacity (as determined from the table), its mass, and its softening length (and the softening kernel). Particle mass and kernel shape are constant in the simulation, opacity is already determined from the table, and the smoothing length is very close to a monotonic function of density. Hence, this new table does not introduce any new independent variables, and is only accurate for particles of a single fixed mass. We calculate the new table by dividing each particle into $2000$ slices of equal column density (but not equal mass). These slices are thin enough that each is optically thin, and we can use the Cloudy tables to determine the dependent properties (temperature, density, heating, cooling) of each slice, because they do not vary much over each slice. We then perform a mass-weighted average over the slices to determine the new table values for gas particles of that density.
  \item We initialize our disk to have a fairly low surface density, to reduce the number of high density particles where self-shielding dominates.
\end{itemize}

\begin{figure}
\begin{center}
\includegraphics[width=\columnwidth]{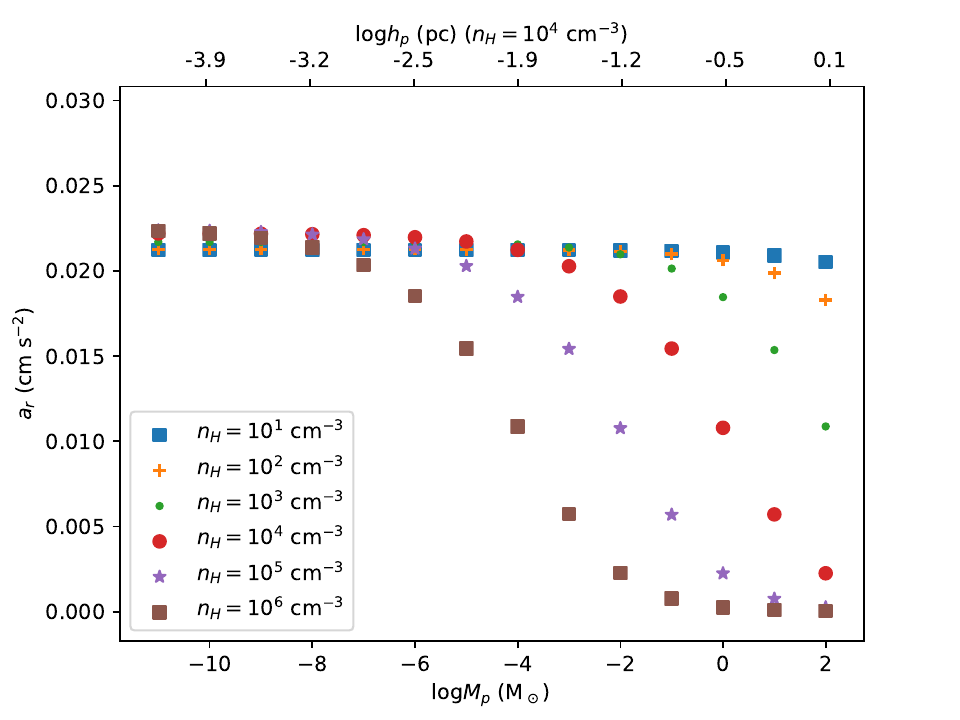}
\end{center}
\caption{\label{hweighted_table}
Radiative accelerations of particles at a fixed temperature ($100$ K) and a fixed AGN flux ($10^6$~erg$^{-1}$s$^{-1}$cm$^{-2}$) as a function of particle mass resolution. The smoothing lengths for particles of density $n_H=10^4$ cm$^{-3}$ are shown on the upper axis.}
\end{figure}

To demonstrate the importance of self-shielding on individual particles, the radiative accelerations particles of temperature $T=100$ K from an AGN flux of $10^6$~erg$^{-1}$s$^{-1}$cm$^{-2}$ of various densities and particle masses are plotted in figure~\ref{hweighted_table}, along with the typical smoothing lengths of particles of this mass resolution with a density of $n_H=10^4$ cm$^{-3}$. At large particle masses and high densities, self-shielding means that the radiation is absorbed in a thin skin and then spread over the mass of the entire particle, suppressing the acceleration. If self-shielding was ignored, and the entire particle received the acceleration that is received at its face, then the radiative acceleration would be massively over-estimated instead. We choose a particle mass of $10^{-4}$ $M_\odot$, but note that even with this fine resolution, acceleration is significantly suppressed for densities greater than $10^5$ cm$^{-3}$. As a basis of comparison with Eulerian simulations, we calculate the typical smoothing length of particles with a mass of $10^{-4}$ $M_\odot$ and a density of $n_H=10^4$ cm$^{-3}$ to be about $10^{-2.5}$ pc. To resolve the radiative acceleration of dense dusty gas, an Eulerian simulation must have a resolution at least this fine.

\begin{figure}
\begin{center}
\includegraphics[width=\columnwidth]{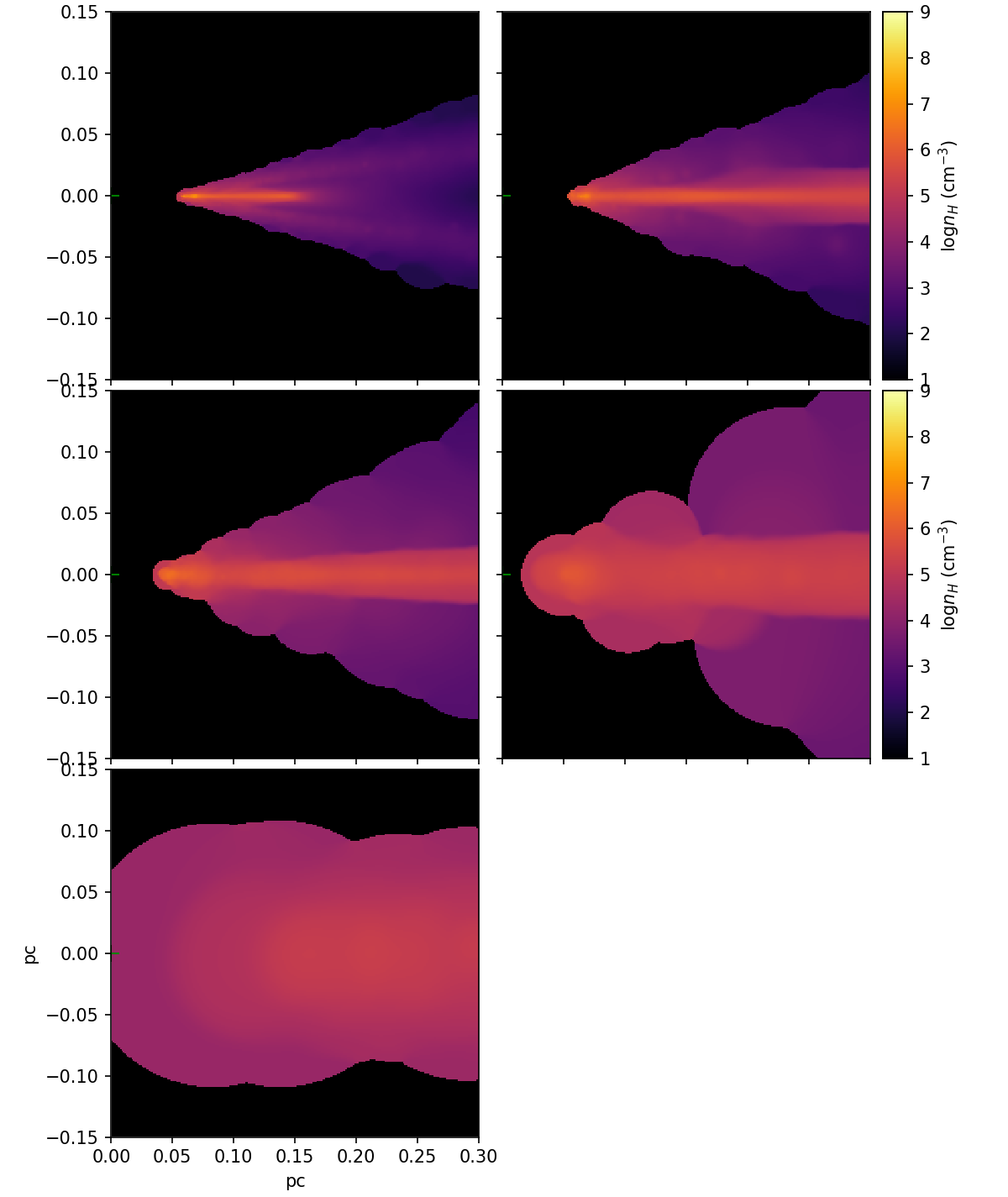}
\end{center}
\caption{\label{resmontage}
Cylindrically averaged gas densities at $t=1.96$ kyr at various particle mass resolutions. Top row: $m_p=10^{-5}$ M$_\odot$,  $m_p=10^{-4}$ M$_\odot$. Middle row: $m_p=10^{-3}$ M$_\odot$, $m_p=10^{-2}$ M$_\odot$. Bottom row: $m_p=10^{-1}$ M$_\odot$.  }
\end{figure}

\begin{figure*}
\begin{center}
\includegraphics[width=\textwidth]{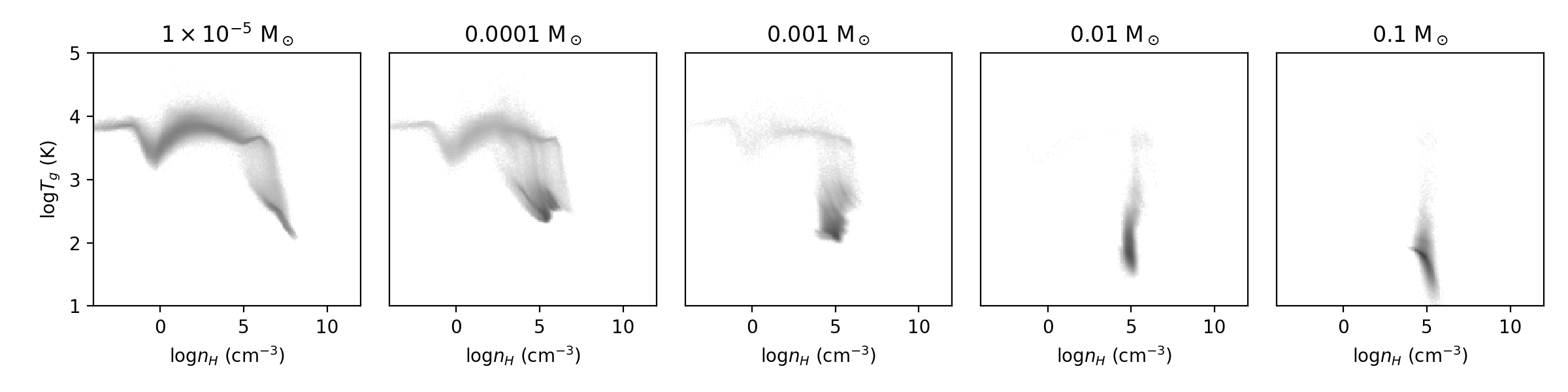}
\end{center}
\caption{\label{resphaseplots}
Gas density-temperature phaseplots as a function of particle mass resolution. }
\end{figure*}

\begin{figure}
\begin{center}
\includegraphics[width=\columnwidth]{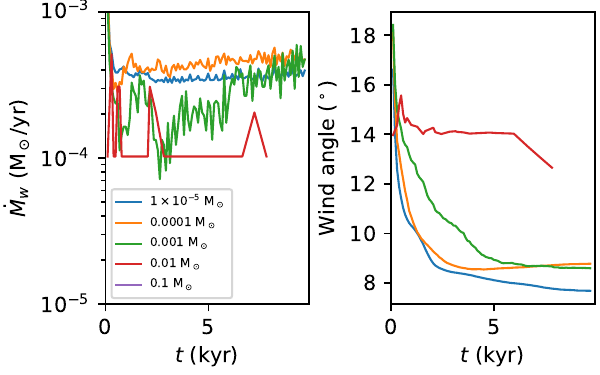}
\end{center}
\caption{\label{restime}
Outflow rate (left), and mean wind inclination (right) for test runs of different resolution across the full simulation time.}
\end{figure}

As an illustration of this, we performed test simulations with resolutions from $10^{-5}$ M$_\odot$ to $10^{-1}$ M$_\odot$. To reduce computational expense in these tests, we kept the total number of particles constant, and the initial surface density constant, and so the outer edge of the initial disk is larger with increasing particle mass. Plots of the density distribution of the gas are in figure~\ref{resmontage}. The biconical wind is only formed at the highest resolutions. This can be seen in the phase-plots in figure~\ref{resphaseplots}, where the hot wind phase is either not present or poorly resolved at $m_p\geq0.001$ M$_\odot$. High density gas is also not resolved at higher resolutions, and much of the gas over-cools. The outflow rates and mean wind inclinations are also dependent on resolution (figure~\ref{restime}). We still see some resolution dependence at our finest test resolution, and so our simulations are not entirely converged with respect to resolution, but we find this dependence weakens beyond a threshold resolution of $\sim10^{-4}$ M$_\odot$. At this resolution, some of the gas still reaches densities high enough that their radiative acceleration would still be theoretically suppressed. However, the resolution dependence becomes weaker, as much of this very dense gas is shielded from the AGN by a large optical depth, and  receives very little radiative acceleration regardless of resolution. Overall, these tests support that a particle resolution coarser than our chosen resolution of $10^{-4}$ M$_\odot$ will not accurately capture the generation of a dusty wind by radiation pressure. 

\subsubsection{Anisotropy of radiation field}\label{anisotopic_flux}

Within the hydrodynamic simulation, the AGN flux is assumed to be anisotropic, following
\begin{equation}
F(r,\theta)=\frac{L}{4\pi r^2} f(\theta)
\end{equation}
where $L$ is the luminosity of the AGN, $r$ is the distance from the AGN, $\theta$ is the angle from the polar axis of the AGN, and the anisotropy function $f(\theta)$ is defined to be
\begin{equation}
f(\theta)=\frac{1+a\cos\theta+2a\cos^2\theta}{1+2a/3}
\end{equation}
where we define $a=(\eta_a-1)/3$, introducing the parameter $\eta_a$ as the ``anisotropy factor'', equal to the ratio between the polar flux and the equatorial flux. This is an generalisation of the classic angle-dependence  of  \citet{1987MNRAS.225...55N}, to account for deviations from perfect limb-darkening of a geometrically thin disk. It is effectively a superposition of an isotropic component with the classic limb-darkened disk. This chosen generalisation produces only a small variation in polar flux with $\eta_a$, despite a large variation in equatorial flux.

The anisotropy factor $\eta_a$ is not well constrained by observations, and strongly depends on the model assumed for the accretion disk and corona. It is likely to vary by wavelength, as emission from the corona should be more isotropic than emission from the accretion disk. In this work, we make the simplification that $\eta_a$ is independent of wavelength, as this allows us to assume a constant SED shape before extinction is included. We also assume that $\eta_a$ is a free parameter, and vary it between the simulations.

Accretion of dusty gas along the disk plane is possible at large $\eta_a$ and low Eddington factors $\gamma_\mathrm{Edd}$, where the equatorial flux is low. However, in test simulations we found the narrow region where dusty accretion is possible was not properly resolved, even at the very fine particle resolution of $m_P=10^{-4}$~M$_\odot$. The result was that radiation pressure would blow away most of the inner surface of the disk \citep[see also][]{2016ApJ...825...67C}, but a few particles would be left behind in the disk plane -- too few to be properly resolved in the SPH kernel. To avoid this problem, we ensure that $\eta_a$ is low enough and $\gamma_\mathrm{Edd}$ high enough that dusty accretion is not possible. Gas, however, can avoid being blown away if the dust is destroyed/sublimated.

\subsubsection{Infrared radiation from the dusty gas}

We are currently developing a model for secondary radiation between dusty gas particles in the simulation, representing the absorption of scattered radiation and dust IR emission by dust. This should provide an additional source of radiation pressure in a vertical rather than radial direction, puffing up the gas and contributing further to the large poloidal extent of the mid-IR emitting material. However, this involves a significant increase in computational complexity and running time, and this model is not yet complete, hence not included in the present simulations. Other groups have included this secondary radiation using the FLD method \citep{2016ApJ...819..115D}, by using the reduced speed of light approximation \citep{2007ApJ...661...52K}, or have excluded it completely \citep{2015ApJ...812...82W} as we do in this current work. In our upcoming paper we will use a ray-tracing method that concentrates on the wind-producing region. It does not require the assumption of optical thickness needed for FLD and does not scale down the system as is implied with a reduced speed-of-light approximation, but does require assumptions about which regions of emission are most important. This secondary radiation should thicken the disk, and to make a very approximate preview of the consequences of this, we assign a temperature floor in some of our runs to introduce an additional source of pressure.

\subsection{Cloudy Models}\label{appendix_cloudy}
We use version c17.00 of the photoionization code Cloudy to incorporate the detailed microphysics that determines the ionization, chemical and thermal states through a slab of dust and gas exposed to an external radiation field.
Our Cloudy models take three parameters as input: the volumetric density of the cloud, the incident AGN radiation field, and the temperature of the gas. The models are in non-local thermal equilibrium, allowing the dust temperature to reach equilibrium with the radiation field and the gas temperature. Each run produces outputs as a function of hydrogen column density, up to $10^{26}$~cm$^{-2}$. 

It has been suggested that the properties of AGN winds do depend on the details of the driving SED \citep{2017MNRAS.467.4161D}, but we focus on a single SED in this work. For the incident radiation field we assume an AGN SED with a Big Blue Bump component at a temperature of $T=10^{5.5}$ K and a logarithmic X-ray to UV flux ratio of $-1.40$. The slope of the Big Blue Bump and the slope of the X-Ray component are left at the Cloudy default values of $-0.5$ and $-1.0$. This produces less inverse Compton scattering of X-ray photons to radio wavelengths than Cloudy's built-in AGN continuum command and is the preferable method as per the Cloudy handbook. We also include the CMB, the cosmic ray background, and a stellar background with an integrated intensity of $1000$ Habing units, which accounts for the nuclear clusters found in the centres of galaxies. This stellar background intensity is taken from \citet{2012ApJ...758...66W}, who note that the exact value does not greatly affect the dynamics of the circumnuclear region \citep{2009ApJ...702...63W}.

We adopt the built-in interstellar medium dust grain model in Cloudy and use the command \texttt{function sublimation} to account for sublimation where applicable. The grain model includes graphites and silicates with sizes following the MRN distribution \citep{1977ApJ...217..425M}, ranging from 0.005 $\mu$m to 0.25 $\mu$m. We do not include PAHs in our calculations as there is an expectation that these very small grains are destroyed close to the AGN \citep[but see][]{2017MNRAS.470.3071J}. We define the dust temperature used in visualisations as the cross-section-weighted average over all dust sizes. For $T>10^5$ K, we assume that the dust has been destroyed by sputtering rather than sublimation, and do not include dust at those high temperatures.

The table covers gas temperatures from $\log{T \mathrm{(K)}}=1-8$, gas densities from $\log{n \mathrm{(cm}^{-3}\mathrm{)}}=0-9$, and AGN fluxes from $\log{I}=1.2-9.2$, where $I$ is in units of erg~cm$^{-2}$~s$^{-1}$. However, for $\log{n \mathrm{(cm}^{-3}\mathrm{)}}>7$ and $\log{T \mathrm{(K)}}>3$, the Cloudy simulations did not converge. Such hot dense gas is rare, and so we extrapolate the table for the few gas particles in this region.

For a given set of the three input values, we compute a total of 8 physical quantities: the heating and cooling rates, the temperature of the dust, the dust to gas mass ratio, the radiative acceleration due to the incident radiation, the absorption and scattering opacities, and the optical depth. The radiative acceleration, intensity-weighted opacities, and optical depth were extracted by calling repeatedly the \texttt{save continuum emissivity} command for each of the $5277$ frequencies considered in Cloudy \footnote{We note that this approach requires a small modification to the Cloudy code to increase the sizes of the relevant storage arrays.} and using a post-processing routine to integrate over the resulting spectrum. The spectrum is sampled in the interval $1.296 \times 10^{-8} - 9.080 \times 10^{6}$ Ryd with nearly logarithmically increasing widths.

\subsection{Self-gravity, star formation and supernovae}\label{section_starform}

Gravitational acceleration is calculated by the sum of an analytic background term (Section~\ref{simulations_section}) and a self-gravity term. Brute-force gravitational acceleration could be calculated through
\begin{equation}
\mathbf{g}_i = \sum_j \frac{Gm_j\mathbf{r}_{ij}}{r_{ij}^3} + \mathbf{g}_\mathrm{bg},
\end{equation}
where $G$ is the universal gravitational constant, $\mathbf{r}_{ij}$ is the displacement between particles $i$ and $j$, and $\mathbf{g}_\mathrm{bg}$ is the analytic background acceleration. A Barnes-Hut oct-tree \citep{1986Natur.324..446B} is used to reduce the size of the sum.

Cold, dense, self-gravitating gas tends to be susceptible to gravitational collapse. In the absence of a feedback mechanism, this collapse continues until stars are formed. The smoothing length of the gas is adaptive, and so we can follow this collapse to arbitrary densities without violating the Truelove criterion \citep{1997ApJ...489L.179T}. However, as we note in Section~\ref{section_rad}, heating, cooling, and radiation pressure must be extrapolated once the density is higher than we have tabulated. As we additionally note, there is also a threshold density beyond which an individual particle's optical depth is greater than unity, and the important skin region is not resolved. 

To prevent this in early tests, the surface density was set so that the Toomre parameter \citep{1964ApJ...139.1217T} was $Q=2$ at all radii, and hence the gravitational instabilities that produce star-forming clumps are suppressed. In these lower resolution tests, we found that radiation pressure nevertheless drove gas to high enough densities that it became Jeans unstable and underwent unimpeded collapse. To prevent this, we implemented a star-formation routine that converts gas above a threshold density into stars at the free-fall time-scale. However, we found that the high resolution of our production runs is sufficient for shear and turbulence to halt the collapse, and star formation never occurs within the simulated disk. This implies that we would not expect significant star formation within the inner few parsec around an AGN, while it may occur at slightly larger distances. 

We have also implemented a supernova feedback model, using a kinetic feedback method, which we use in some of our simulations. Our simulation time is shorter than the life times of even the most massive stars, and so we cannot self-consistently calculate the rate and location of supernovae based on resolved star formation. Instead, we place supernovae at a fixed rate in an evenly weighted random location in the disk within a distance $R_\mathrm{SN}$ of the AGN center, where $R_\mathrm{SN}=0.4$~pc is a model parameter. The kinetic energy of a supernova is $\epsilon_\mathrm{SN}E_\mathrm{SN}$, where $E_\mathrm{SN}=10^{51}$ erg is the fiducial energy output of a supernovae, and $\epsilon_\mathrm{SN}$ is the fraction of energy that is converted into kinetic energy and is a model parameter, which we set to $\epsilon_\mathrm{SN}=0.1$. Supernovae are assumed to occur instantaneously, and a supernova event affects all $n$ particles within $r_\mathrm{SN}$ by giving each particle a radially directed kinetic energy kick $\epsilon_\mathrm{SN}E_\mathrm{SN}/n$, where $r_\mathrm{SN}=0.01$~pc is a model parameter.

\subsection{Simulations}\label{simulations_section}

\begin{table}
\begin{center}
\begin{tabular}{lccccc}
\hline\hline
~               & ~                     & ~        & $T_\mathrm{floor}$ & SNR        & $F_p/\langle F\rangle$\\
Name            & $\gamma_\mathrm{Edd}$ & $\eta_a$ & (K)               & Myr$^{-1}$ & ~ \\
\hline
NoAGN            &  N/A    &         N/A&               $10^1$ & $0$     &  N/A   \\
a2\_e01           & $0.01$ &        $10^{2}$ &        $10^{1}$ & $0$      & $4.35$\\
a2\_e02           & $0.02$ &        $10^{2}$ &        $10^{1}$ & $0$      & $4.35$\\
a2\_e05           & $0.05$ &        $10^{2}$ &        $10^{1}$ & $0$      & $4.35$\\
a2\_e1            & $0.10$ &        $10^{2}$ &        $10^{1}$ & $0$      & $4.35$\\
a2\_e2            & $0.20$ &        $10^{2}$ &        $10^{1}$ & $0$      & $4.35$\\
a2\_e01\_SN100    & $0.01$ &        $10^{2}$ &        $10^{1}$ & $10^{2}$ & $4.35$\\
a2\_e01\_SN1000   & $0.01$ &        $10^{2}$ &        $10^{1}$ & $10^{3}$ & $4.35$\\
a2\_e01\_T30      & $0.01$ &        $10^{2}$ & $3\times10^{1}$ & $0$      & $4.35$\\
a2\_e01\_T300     & $0.01$ &        $10^{2}$ & $3\times10^{2}$ & $0$      & $4.35$\\
a2\_e01\_T1000    & $0.01$ &        $10^{2}$ &        $10^{3}$ & $0$      & $4.35$\\
a0\_e1            & $0.10$ &        $10^{0}$ &        $10^{1}$ & $0$      & $1.00$\\
a1\_e1            & $0.10$ &        $10^{1}$ &        $10^{1}$ & $0$      & $3.33$\\
a3\_e1            & $0.10$ &        $10^{3}$ &        $10^{1}$ & $0$      & $4.49$\\
\hline
\end{tabular}
\end{center}
\caption{\label{ictable} \textup{
Summary of run parameters: $\gamma_\mathrm{Edd}$ is the Eddington factor, $\eta_a$ is the anisotropy factor (i.e. the ratio between the polar flux and the equatorial flux), $T_\mathrm{floor}$ is the temperature floor, and $F_p/
\langle F\rangle$ is the ratio between the polar flux and the isotropic average flux.}
}
\end{table}

We model the initial conditions for the gas in the inner region of the torus as a disk of mass $M=69.0294$ M$_\odot$ and thickness $0.02$ pc, with an inner radius of $0.005$ pc, and an outer radius of $0.47$ pc. The surface density is set to $100$~M$_\odot$pc$^{-2}$, giving an initial density of $\rho=5000$~M$_\odot$~pc$^{-3}$ ($n_\mathrm{H}\approx1.7\times10^5$ cm$^{-3}$). The gas disk has a uniform initial temperature of $1000$ K, but heating and cooling are sufficiently rapid in our simulations that the evolution is not sensitive to the initial temperature. At our high resolution we can not model the entire dusty region out to several 10s of parsecs, but instead concentrate on examining the inner region in detail. For comparison with a full-scale `torus', if the initial surface density of the disk was extended to a radius of $20$ pc, it would have a total mass of $130,000$ M$_\odot$. The gas is modelled by 690,294 particles, each with a mass of $10^{-4}$ M$_\odot$. 

The imposed gravitational potential is a superposition of a softened Keplerian term to represent the black hole, and a Hernquist bulge term to represent the potential of the galactic stars and dark matter. In practice, the Keplerian term dominates for most particle orbits. The form of the gravitational force law is

\begin{equation}
F(r)/G=\frac{M_\mathrm{BH}r^2}{r^2+c_{BH}^2} + \frac{M_\mathrm{H} (r/c_{H})^2}{(1+r/c_{H})^2}
\end{equation}
where $M_\mathrm{BH}$ and $M_\mathrm{H}$ are the masses of the black hole and Hernquist bulge, $c_{BH}$ is the softening length for the black hole potential, and $c_{H}$ is the scale length for the Hernquist bulge. We keep most of the parameters constant in this suite of simulations, setting $M_\mathrm{H}=10^9$ M$_\odot$, $c_{BH}=10^{-4}$ pc, and $c_H=250$ pc, but vary the black hole mass with $\gamma_\mathrm{Edd}$ so that the AGN luminosity is constant at $1.26\times10^{42}$ erg~s$^{-1}$. We set $M_\mathrm{BH}=10^6\mathrm{~M}_\odot (\gamma_\mathrm{Edd}/0.02)^{-1}$. The particles are given velocities such that the initial rotation curve is in equilibrium with the potential, although the thickness of the disk is primarily supported by pressure. Under strong radiation pressure, this means the gas disk is not in equilibrium, but is pushed outwards. Hence when radiation pressure is strong, the simulations do not represent a long-term equilibrium, but rather the short-term response of a gas disk to the AGN radiation field.

We perform multiple simulations from $\gamma_\mathrm{Edd}=0.02$ to $\gamma_\mathrm{Edd}=0.2$. We set $\eta_a=10^2$ in most of the simulations, except for the simulations with $\gamma_\mathrm{Edd}=0.1$, where we also perform simulations with $\eta_a=10^0$ (i.e. isotropic), $\eta_a=10^1$, $\eta_a=10^3$. Most simulations have a temperature floor of $10$ K (which is never reached in the simulation), but in some we introduce higher temperature floors of $T=30,300,3000$ K to see the effects of a thicker disk, as we might expect to be produced by the disk puffing itself up with infrared radiation pressure. We also produce one simulation (`NoAgn') with no AGN radiation field, where gas is evolved adiabatically. We run each simulation for $9.78$ kyr, and all systems either reach a steady state or are blowing out the gas by the end of the simulation.

We use a naming scheme that represents the radiation anisotropy and Eddington factor for each run, as well as the presence of supernova feedback or an increased temperature floor. These simulation parameters are summarized in Table~\ref{ictable}.

\section{Results}\label{section_results}

\subsection{Simulation Evolution}\label{section_evolve}

\begin{figure*}
\begin{center}
\includegraphics[width=\textwidth]{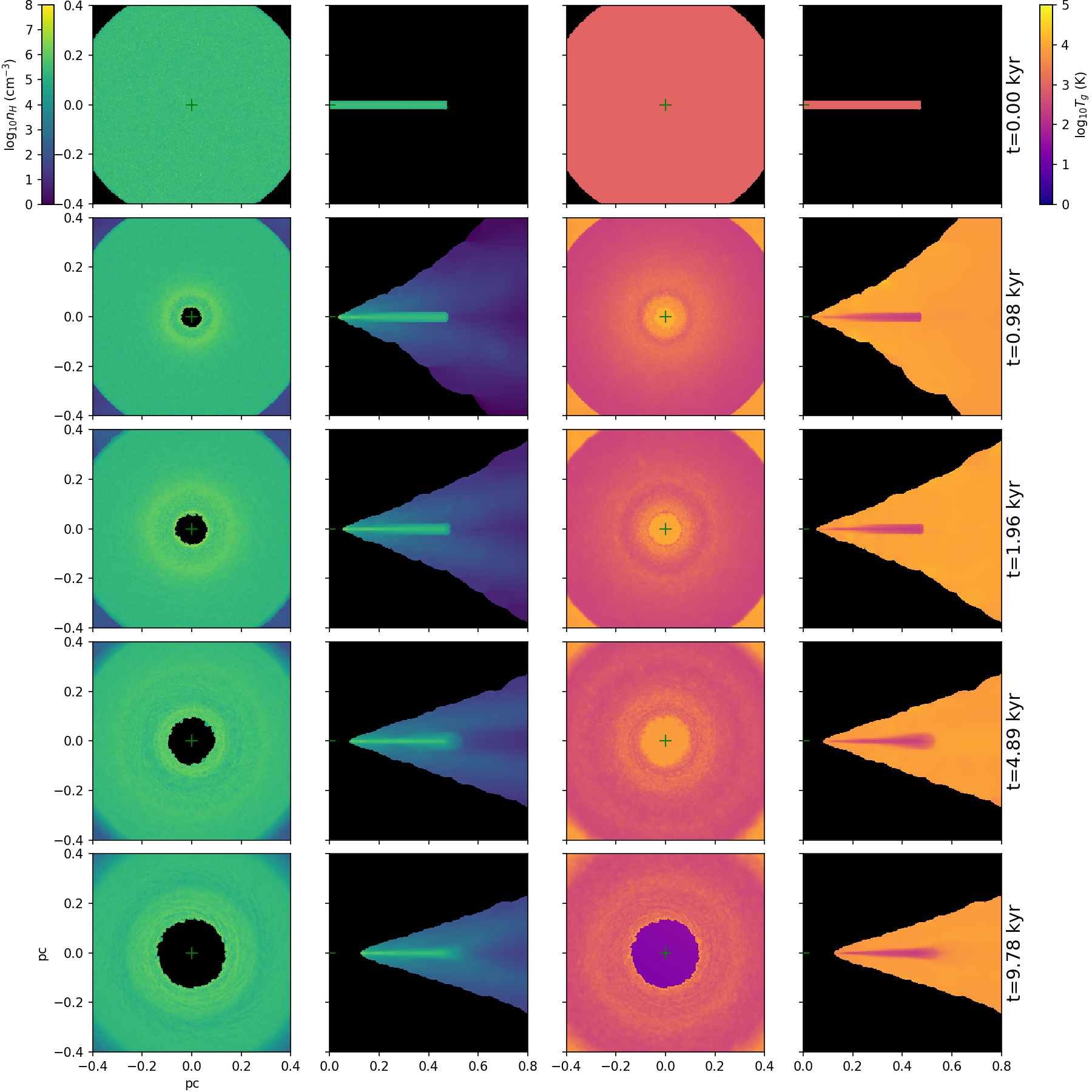}
\end{center}
\caption{\label{sample_evolve}
Evolution of Run a2\_e01. Left: face-on and edge-on mass-weighted mean densities. Right: face-on and edge-on mass-weighted mean temperatures.
}
\end{figure*}

\begin{figure}
\begin{center}
\includegraphics[width=\columnwidth]{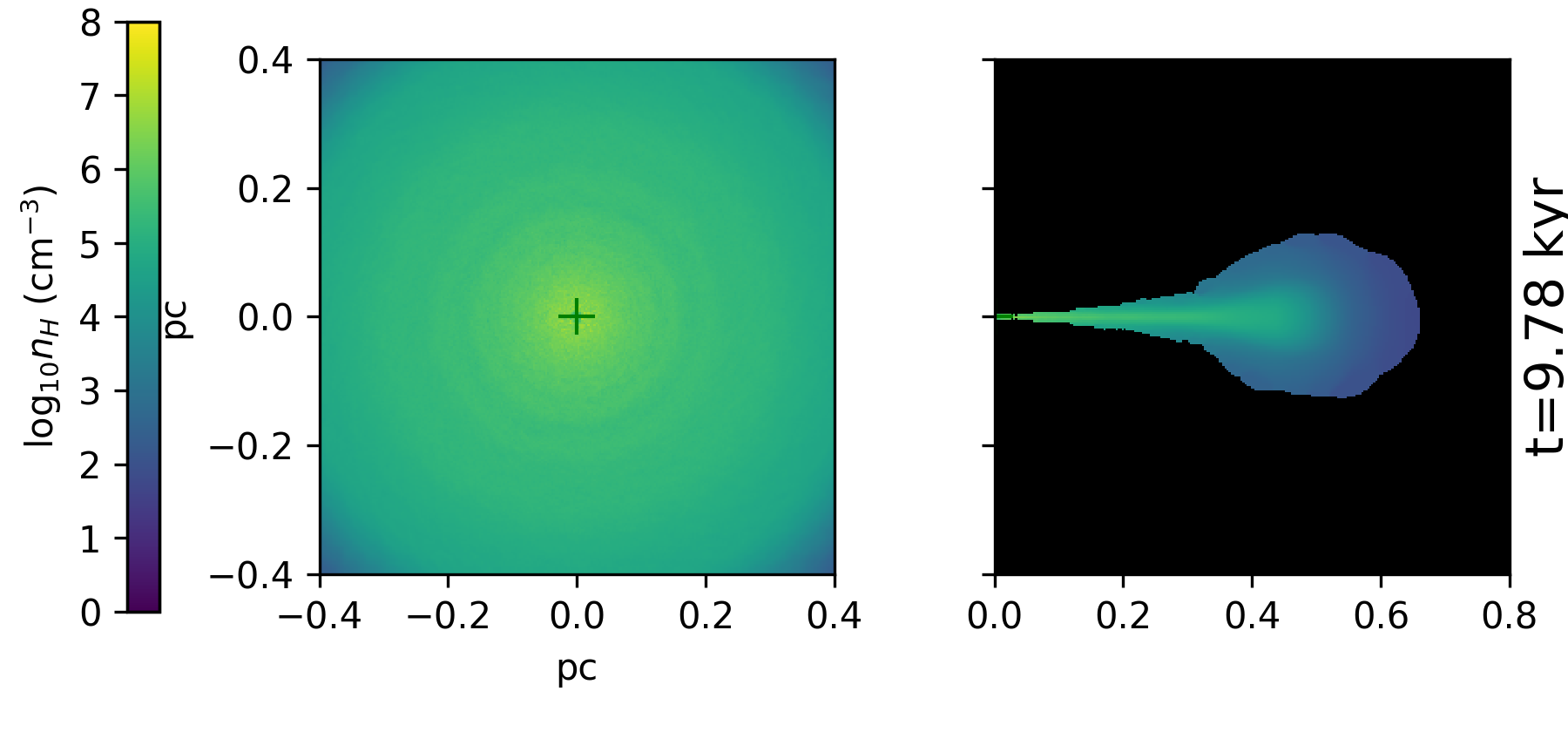}
\end{center}
\caption{\label{noagn_end}
Face-on and edge-on mass-weighted mean densities for Run NoAGN at the end of the simulation.
}
\end{figure}

\begin{figure*}
\begin{center}
\includegraphics[width=1.\textwidth]{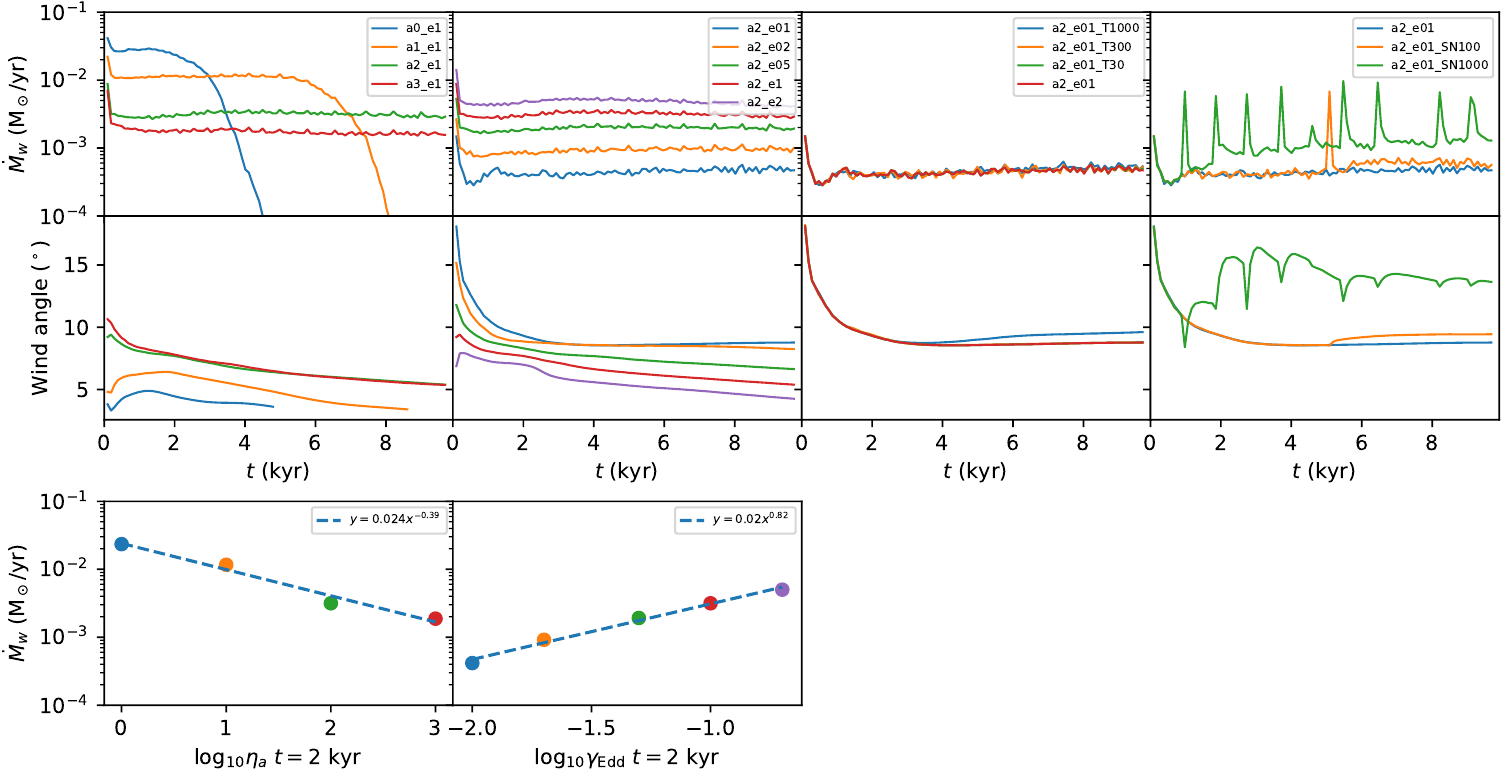}
\end{center}
\caption{\label{timeseries}
Outflow rate (top), and mean wind inclination (centre) for all runs across the full simulation time, and outflow rates at $t=2$ kyr as a function of the varied parameter. From left to right: runs with varying $\gamma_\mathrm{Edd}$; runs with varying $\eta_a$; runs with varying temperature floors; runs with varying supernova rates. Some runs are plotted multiple times to facilitate comparison.}
\end{figure*}

All of the AGN runs follow a similar evolution, and so here we focus on the properties and evolution of Run a2\_01, and only describe the other runs when variations appear. The evolution of the gas density and temperature for Run a2\_e01 are plotted in Figure~\ref{sample_evolve}. The gas disk remains thin and cool, because the heating from the AGN is deposited only in a thin skin at the inner edge of the disk, and is not re-radiated inwards. Gravitational instabilities are visible as ring and spiral patterns, but these do not greatly thicken the disk. In the absence of radiation pressure, and of non-adiabatic heating and cooling, the gravitational potential causes the system to evolve into a flared disk (Figure~\ref{noagn_end}) with no outflow.

A hot wind is formed. The thin skin region on the inner edge of the disk receives strong radiation pressure in addition to heating, causing the inner gas to be pushed outwards as a warm wind. The thermal pressure of the gas pushes it vertically, thickening the outflow. Although the gas of this wind reaches temperatures of $\sim10^4$ K, the dust component is significantly cooler, and is not destroyed by sublimation or sputtering. The opacity of the gas therefore remains large, and the gas continues to be pushed outwards. We expect that the inclusion of re-radiation of infrared radiation between gas particles would provide an additional vertical force through radiation pressure, and will explore this in a future paper.

The inner radius of the disk slowly expands throughout the simulation. This is partially a result of the inner edge of the disk being ablated through heating and radiation pressure, but is also a result of the net momentum deposition on the disk as a whole from radiation pressure. Most of the deposited momentum is carried off in the wind, but some is transferred to the bulk of the disk, pushing the disk outwards. This effect is stronger when the intensity of radiation in the plane of the disk is stronger, especially in Run a0\_e1. The rotation curve of the gas disk is initially in equilibrium with the gravitational potential, and does not represent a self-consistent long-term state of the system, but rather the short-term response of a disk exposed to AGN radiation. We will explore more self-consistent models with gas accretion in a later paper.

\subsection{The emerging wind}

To characterize the wind, we must precisely define which gas particles are considered to be ``wind particles''. In this work, we define the wind as all gas particles with velocities greater than the escape velocity from the gravitational potential at their position, within $5$ pc of the simulation center. This cut-off radius is used to prevent gas expelled early in the simulation from dominating the results. We define the outflow rate $\dot{M}_w$ as the rate at which particles join the wind, plotted in the top row of Figure~\ref{timeseries}.

Considering only the runs with $\eta_a=10^2$, the runs with a greater Eddington factor show a higher outflow rate, quickly reaching a steady state. This suggests a fairly direct relationship between Eddington factor and outflow rate, and plot the outflow rate at $t=2$ kyr against Eddington factor in the bottom row of Figure~\ref{timeseries}. Here, we find a power law relationship between outflow rate and Eddington factor that is slightly sublinear, with a power index of $\sim0.8$.

More critically, we find that varying the anisotropy of the radiation field has a dramatic effect. It has sometimes been assumed that no radiation at all is emitted in the AGN plane, for at least part of the SED \citep[such as in the UV, e.g.][]{2016ApJ...828L..19W}, or conversely that the radiation field is anisotropic across the entire SED \citep[e.g.][]{2007ApJ...661...52K,2016ApJ...819..115D}. We should expect limb-darkening for radiation emitted by the AGN accretion disk, but we should not expect the radiation in the plane to be absolutely zero. The disk is not likely to be perfectly flat and perfectly thin, and we should expect a contribution of reflected radiation from the corona as well. We have varied the equatorial emission from our fiducial value of $1\%$ of the polar emission (the a2\_e* runs) to $0.1\%$ in Run a3\_e1 up to $100\%$ in Run a0\_e1. We find a strong dependence of outflow rate on anisotropy. This is particularly dramatic in a1\_e1 and a0\_e1. Here, the more isotropic radiation field produces a very large radiation pressure on the inner surface of the disk, producing a strong outflow that blows away the gas disk entirely. In the bottom row of Figure~\ref{timeseries} we plot the anisotropy factor against outflow rate, and find a power law trend with an index of $-\sim0.4$. This demonstrates that decisions on the anisotropy of the AGN radiation field can not be taken lightly, as sensitivity of the outflow rate to the anisotropy is of a similar order to its sensitivity to the Eddington factor.

\subsection{Dependence of the wind angle on AGN radiation field and pressure}

We can see related effects in the evolution of the wind angle, which is defined as the mean latitude of all wind particles, plotted in the middle row of Figure.~\ref{timeseries}. The wind angle is higher when the initial outflow rate is lower. This is caused by the interaction between thermal pressure causing the wind to expand above and below the disk, and radiation pressure pushing the gas outwards. The thermal pressure of the wind particles and the disk does not vary greatly between the runs, despite the disk receiving different fluxes. This is the result of negative feedback from radiative cooling that causes the gas to be attracted towards discrete phases of temperature and density -- it requires a large amount of energy to push the gas beyond $T=10^4$ K. However, there is no feedback effect to reduce radiation pressure until dust is destroyed, and in this regime radiative acceleration essentially increases in proportion to the flux on the inner disk edge. Thus at lower anisotropies, the gas receives greater radiation pressure but still has a similar thermal pressure, and the wind is pushed outwards more radially, producing a lower wind angle. Similarly, at higher Eddington factors, the gravitational potential is stronger while the radiation pressure is unchanged, and the outflows also become more equatorial.

The interaction between thermal and radiation pressure is also demonstrated in the simulations with varying temperature floor, in the right panel of Figure~\ref{timeseries}. As a reminder, we use the temperature floor as a means to approximate the effect of radiation pressure from the ambient radiation field within the dusty gas at puffing the disk, that we aim at modelling directly in the upcoming paper. As the wind-launching gas is rapidly heated, the wind is not very sensitive to the previous temperature of the gas. At $T_\mathrm{floor}=1000$ K we do see an increase in the wind angle, due to thermal pressure producing a slightly thicker disk, but this effect is small.

Supernovae also only have a relatively small effect. Each supernovae is visible as a significant but short-lived peak in the outflow rate. In a2\_e01\_SN1000 the supernova rate is high enough that the disturbance caused by supernovae also causes the outflow rate to somewhat increase in-between supernovae, although the increase in mass outflow is smaller than that caused by lowering the anisotropy. The wind angle also becomes significantly more polar, although as we note in Section~\ref{observables}, this represents a small amount of gas thrown to high angles rather than a significant puffing up of the wind in general. 

In these small scale simulations, the coupled energy input of a supernova ($10^{50}$ erg, see Section~\ref{section_starform}) is greater than the initial kinetic energy of the disk ($\sim10^{49}$ erg). Supernovae therefore have an extremely small effect compared to their huge energy input. This is because the energy input not efficiently coupled to the disk as a whole. Instead, as the energy deposition region is well resolved, it forms a bubble of hot gas and immediately escapes in a ``chimney''. Rather than boosting the wind and disk in general, a small amount of hot gas is ejected at high speeds.

We also note that, assuming 1 SN per $\sim100$~M$_\odot$ of star formation, the rate of $1000$~SNe/Myr is the equivalent of a star formation rate of about $0.1$ M$_\odot$~yr$^{-1}$ concentrated within a $\sim0.47$ pc disk, or a star formation surface density of $>4\times10^5$ M$_\odot$~yr$^{-1}$~kpc$^{-2}$. This modest effect on the disk and wind is therefore the result of an extremely powerful nuclear starburst that we should not expect to be ubiquitous in obscured active galactic nuclei.

\subsection{Observable Properties}\label{observables}

\begin{figure*}
\begin{center}
\includegraphics[width=\textwidth]{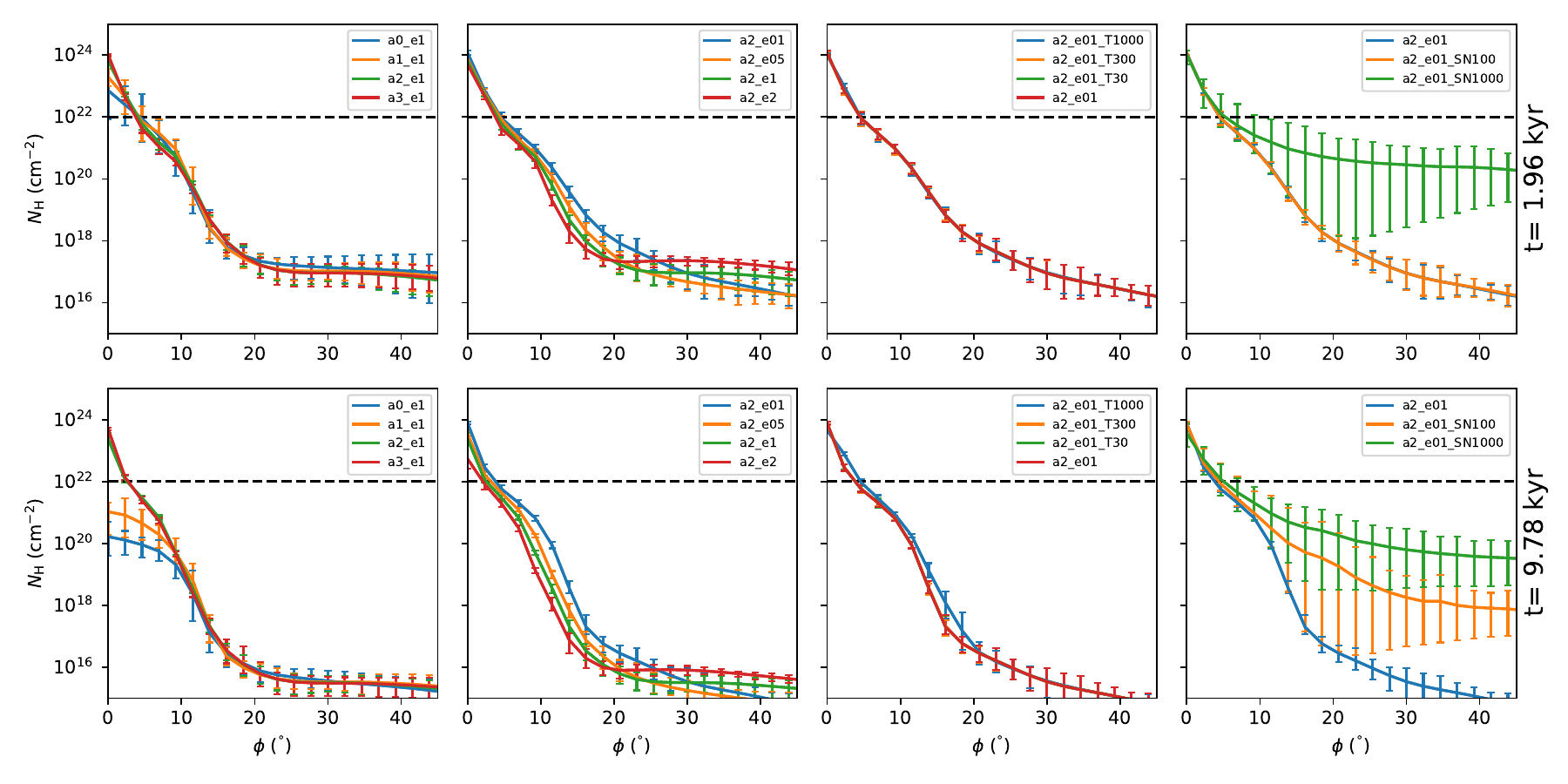}
\end{center}
\caption{\label{sigma_angle}
Mean column densities at $t=1.96$ kyr and $t=9.78$ kyr as a function of inclination from the disk plane, $\phi$. The error-bars are the maximum and minimum of the azimuthal variation. Runs are grouped as in Figure~\ref{timeseries}.
}
\end{figure*}

\begin{figure}
\includegraphics[width=\columnwidth]{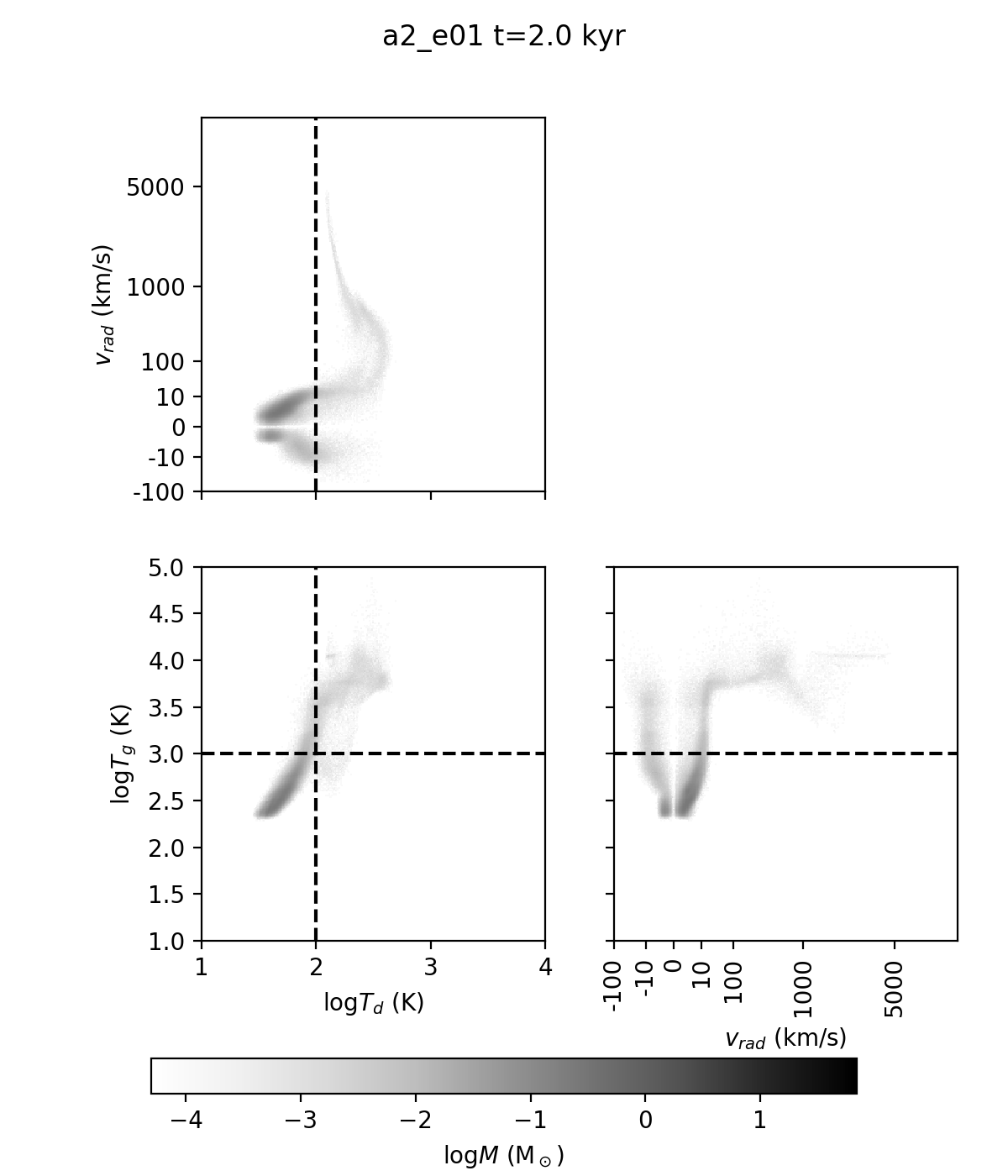}
\caption{\label{tv_phaseplots}
Phase-plots of the gas temperature ($T_g$), dust temperature ($T_d$), and radial velocity ($v_{rad}$) of Run a2\_e01 at $t=1.96$ kyr. The dotted line indicate the gas and dust temperature cuts used in Figure~\ref{streamlines}.
}
\end{figure}

\begin{figure}
\begin{center}
\includegraphics[width=\columnwidth]{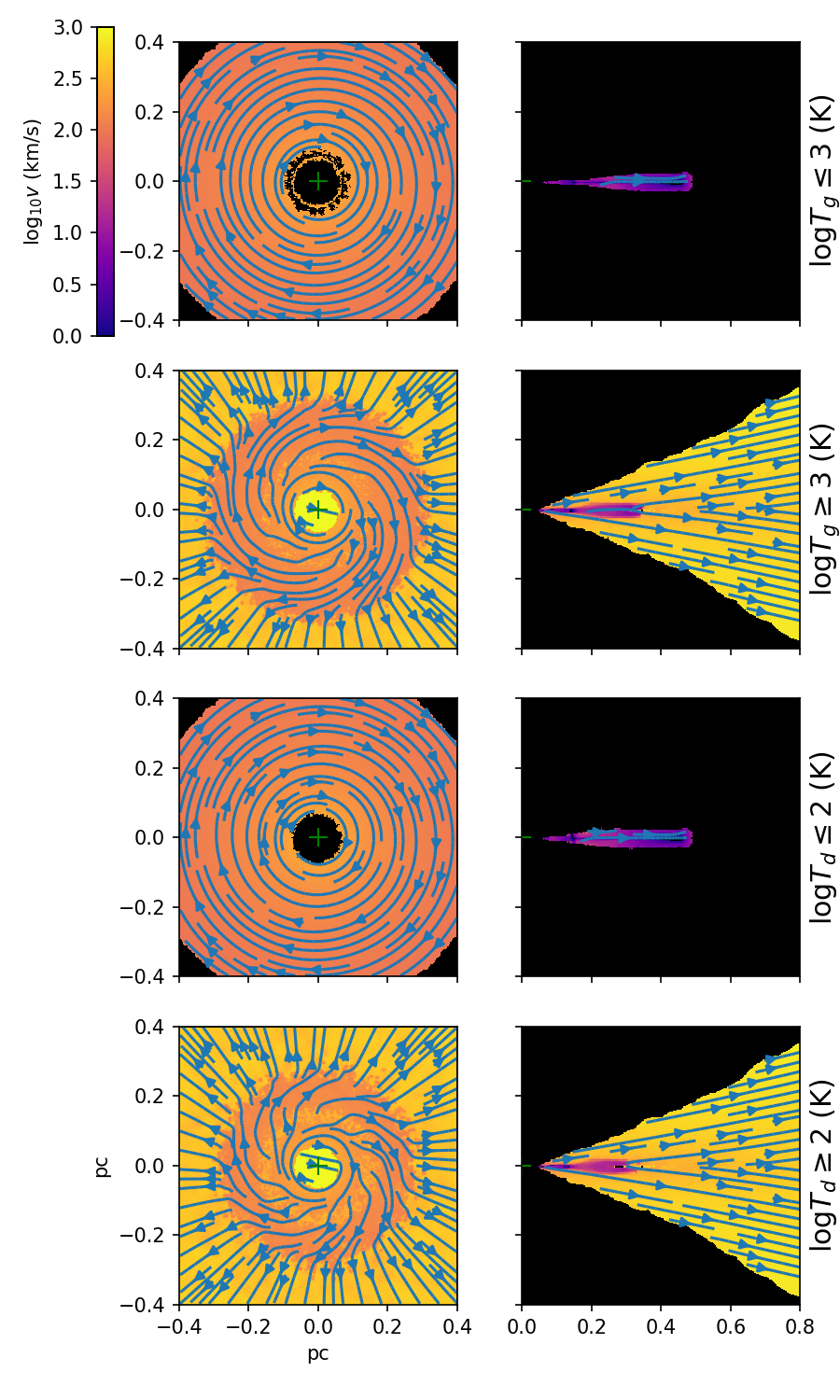}
\end{center}
\caption{\label{streamlines}
Mass-weighted mean velocities of dust and gas temperature above and below threshold temperatures, for Run a2\_e01 at $t=1.96$ kyr. Left column: face-on view. Right-column: azimuthally wrapped view. The color-map gives the magnitude of the velocity, and the streamlines give the direction.
}
\end{figure}

\begin{figure*}
\begin{center}
\includegraphics[width=\textwidth]{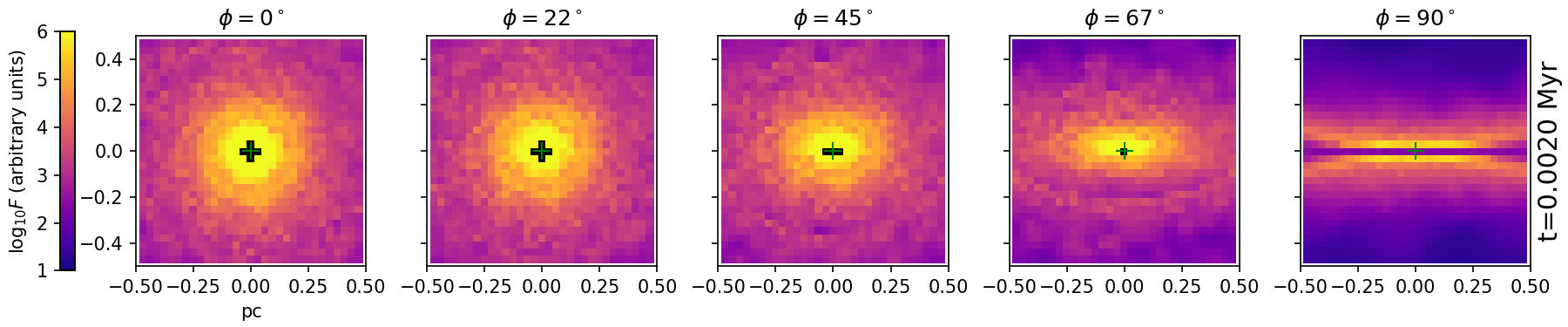}\\
\end{center}
\caption{\label{rayview}
Raytraced mock-images of Run a2\_e01 at $t=1.96$ kyr.
}
\end{figure*}
\begin{figure*}
\begin{center}
\includegraphics[width=\textwidth]{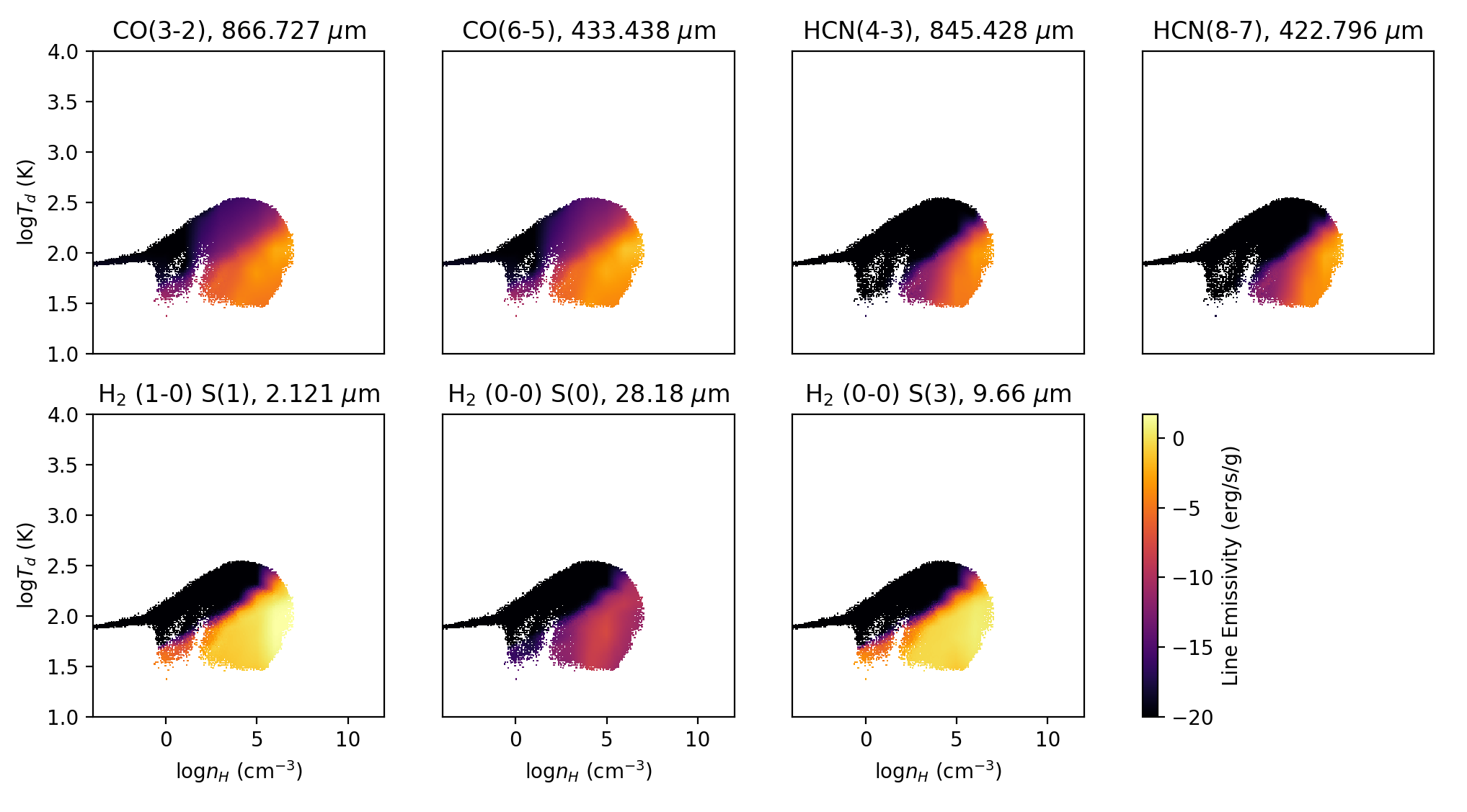}\\
\end{center}
\caption{\label{linephases}
Line intensities as a function of the phases of dust temperature and gas density for Run a2\_e01 at $t=9.78$ kyr.
}
\end{figure*}

\begin{figure*}
\begin{center}
\includegraphics[width=\textwidth]{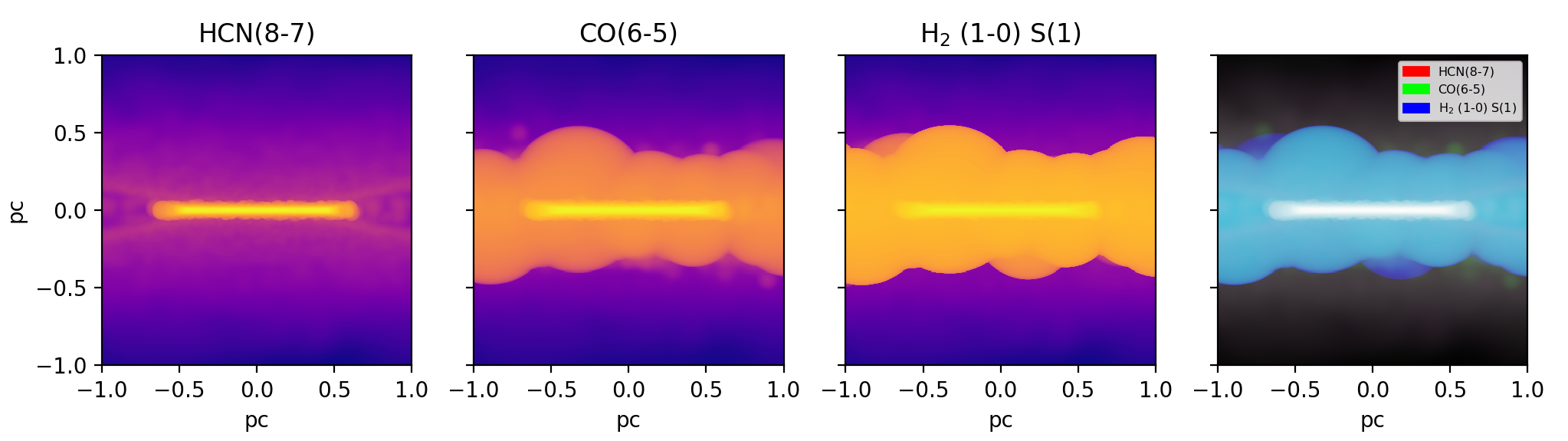}\\
\end{center}
\caption{\label{lineimages}
Line images for Run a2\_e01 at $t=9.78$ kyr.
}
\end{figure*}

\subsubsection{Column Density Profiles}

AGN unification requires column densities with a large covering fraction, and so the column density as a function of angle above the disk plane is a critical constraint for AGN torus models. Although it is not possible to observe a single AGN from multiple angles, the general shape of the column density distribution can be inferred statistically from observing a large number of AGN.

We calculate the column density for each angle by tracing rays from the centre of the simulation outwards, and summing the contribution to column density across all particles that intersect the ray. We convert the mass surface density into a number surface density by assuming a mean molecular mass of $\approx1.22$. For each of $40$ different inclinations, we trace $40$ rays, equally spaced azimuthally. The mean, maximum, and minimum of each set of rays for all runs at $t=1.96$ kyr and $t=9.78$ kyr are plotted in Figure~\ref{sigma_angle}. We plot the earlier time because the gas disk is blown away at later times in some runs, and the column densities become very small, and we plot the later time because the effects of supernovae do not become apparent until later in the simulation.

The simulations all have remarkably similar column density profiles, but there are some trends to be observed. Runs a0\_e1 and a1\_e1 have less material at very low inclinations as the disk is blown away. We see a related effect when comparing runs with different Eddington ratio. Here the column density slowly decreases with increasing Eddington ratio, except at $\phi>20^\circ$, where the trend is perhaps reversed, although this represents only a very small amount of gas.

As above, the temperature floor has little effect. This indicates that the details of the hydrodynamics of the disk are not important for the wind - in the wind generation region, the gas properties are dominated by the radiation of the AGN and the intervening opacity, and small changes in disk thickness are unimportant.

Supernovae do have an effect at large inclinations, propelling small amounts of gas and dust to large heights about the disk. The effect here is not to greatly change the opening angle of the optically thick part of the wind, but to provide a small level of baseline extinction at more polar viewing angles.

\subsubsection{Kinematics}

Specific observational tracers are typically sensitive to gas or dust at some particular range of temperatures. To demonstrates what dynamical phases a particular observation might capture, phase-plots of gas temperature, dust temperature, and radial velocity from the simulation centre of Run a2\_e01 at $t=1.96$ kyr are plotted in Figure~\ref{tv_phaseplots}. The gas mostly lies in two distinct phases, with some gas in intermediate states - wind gas in the process of being heated. The bulk of the gas mass remains in the disk, with a radial velocity of $|v_{rad}|\lesssim10$ km/s, a gas temperature of 100s of K, and a dust temperature of 10s of K. Most of the rest of the gas mass is either transitioning into wind or already in the wind - a warm ionized phase with gas temperatures of just under $10^4$ K.

The disk phase is clearly visible in the phase-plots in Figure~\ref{tv_phaseplots} as having low dust and gas temperatures. We can isolate this phase by applying cuts of either gas or dust temperature, as plotted in Figure~\ref{streamlines}. We choose $T_d=100$ K and $T_g=1000$ K as threshold temperatures, and plot the velocity fields of gas above and below these thresholds in Figure~\ref{streamlines}. Both the gas and dust temperature cuts select qualitatively similar velocity fields, which is expected as there is a correlation between dust and gas temperature, at low gas temperature. Material with temperature below either cut shows strong rotation, while material above either cut is strongly outflowing. This suggests that observational tracers of hot/cold gas will respectively trace hot/cold dust and vice-versa, and that tracers of hot material should observe outflowing gas, and of cold material should observe indicators of rotation. However, as we note below, this coupling is not exact.

\subsubsection{Mock images}

In Figure~\ref{rayview}, we have plotted ray-traced mock images of Run a2\_e01 from various angles. For each ray - corresponding to a pixel in each image - intersecting particles are sorted by distance along the line of sight. Summing from nearest to furthest, each particle contributes a flux equal to $\sigma_B f_d T_d^4 e^{-\tau}$, where $\sigma_B$ is the Stefan-Boltzmann constant, $f_d$ and $T_d$ are the dust mass fraction and dust temperature, as calculated with the Cloudy tables, and $\tau$ is the optical depth from all previous particles along the line of sigh. He we do not use the Cloudy table opacities, as these are weighted by the flux of the AGN. Instead for this visualization, we use an opacity of $65.2$ cm$^{2}$g$^{-1}$, which is typical for ISM dust. The normalization of the plotted flux is arbitrary.

The model images reproduce some features that are observed in infrared interferometry \citep{2012ApJ...755..149H,2013ApJ...771...87H}. We find a bright ring of hot dust in the centre of the disk, as observed in near-IR. In the edge-on view, this is no longer visible, but is blocked by extinction from the disk, and most of the emission appears above and below the disk, representing the warm wind. This warm wind corresponds to the polar extended mid-IR emission, although in our simulations the wind does not extend as far in the polar direction as interferometry suggests due to the lack of IR-reradiation.

Recently, ALMA observations have produced high resolution imagery of nearby active galactic nuclei on slightly larger than parsec scales \citep[e.g.][]{2016ApJ...823L..12G,2017A&A...608A..56G,2018ApJ...859..144A,2018ApJ...853L..25I,2018ApJ...867...48I}, observing molecular lines from molecules such as HCN and CO. Using CLOUDY, we have produced tables of emissivities for two CO lines, two HCN lines, and three vibrational transitions of H$_2$. The CO and HCN lines lie within the range observable by ALMA, while the H$_2$ lines are observable by the VLT and VLTI instruments (H$_2$ 1-0 S(1) and H$_2$ 0-0 S(3)), or in the future with JWST ((H$_2$ 1-0 S(0), although only at low redshift). The emissivities as a function of dust temperature and gas density are plotted in Figure~\ref{linephases} for Run a2\_e01 at $t=9.78$ kyr, the end of the simulation. Generally, the lines are stronger at high gas densities and low dust temperatures, with a rapid transition from strong emissivity to zero emissivity as the molecules are destroyed in hotter, lower density environments. For these lines, dust temperature is a proxy for the intensity of radiation. We note above that there is a correlation between dust temperature and gas temperature, but the molecular emission is more closely tied to dust temperature than to the gas temperature, and the emissivity cutoff is less clean in a gas temperature-density plot.

All of the lines trace high density gas with cool dust, but differ in total strength, and in the slope and location of the transition to non-emitting. The CO lines, and the H$_2$ (1-0) S(1) and H$_2$ (0-0) S(3) vibrational lines retain emission in higher temperatures and lower densities than the HCN lines, with the $H_2$ (0-0) S(0) line lying somewhere in-between. To show how this would might appear in high resolution observations, we have plotted the fluxes from three of these lines in Figure~\ref{lineimages}, along with a combined RGB image. The flux in each pixel is simply sum of the column density multiplied by the tabulated emissivity per particle within one smoothing length of the line of sight through the pixel, and does not include extinction effects. Here it can be see that the HCN lines trace material closest to the disk, including part of the outflowing wind. The CO line trace material further from the disk, and the H$_2$ 1-0 S(1) vibrational line emphasizes the extended material even more. The expectation from these simulations is that the H$_2$ 1-0 S(1) lines shows a higher velocity dispersion than the HCN and CO lines, which is consistent with observations of these lines \citep[e.g.][and references therein]{2009ApJ...696..448H}.

\section{Discussion}\label{section_discussion}

In this work, we have concentrated on disk winds as a critical component of the observed ``torus''. In this work we have neglected IR radiation pressure that could potentially puff up the disk into a toroidal shape, although we will incorporate this in a future paper. However, it is unclear whether a radiation-pressure supported torus is a sufficiently robust and stable model to explain the geometrically thick obscuration implied from observations, and that a wind-based model may be more universal. The simulations of \citet{2016MNRAS.460..980N} were performed on a simulation length scale and resolution to our simulations, and included IR radiation pressure, but did not find the disk is vertically supported to form a torus. They point out that the thick toruses produced in previous work occur in simulations with assumptions and parameters that over-estimate the role of radiation pressure support, such as assuming that the gas temperature is equal to the dust temperature \citep{2016ApJ...825...67C}, or assuming a larger x-ray component to the SED along with flux-limited diffusion \citep{2016ApJ...819..115D}. However, we note that \citet{2016ApJ...825...67C} do not find that the `torus' is fully supported against gravity at low UV luminosities, and emphasize the presence of a wind in their models. It does appear that the wind is the future of AGN obscuration models, as we emphasize in this work. 

The dramatic effect of varying the anisotropy of the radiation field, at constant luminosity and Eddington factor, demonstrates that this is a parameter that can not be chosen arbitrarily. The expected anisotropy depends on models of the accretion disk and corona, and we should not expect the radiation field to be purely anisotropic or purely fit a thin-disk model. Although much work has gone into developing sophisticated radiation transfer schemes, the correct choice (or choices) for this single parameter may have a more dominant effect. The level of anisotropy can be constrained by simulations of the accretion disk and corona emission, but these results are still model dependent \citep{2015MNRAS.449..191X}, and variations in this parameter need to be explored.

Isotropic radiation was applied in the simulations of \citet{2016ApJ...825...67C}, and it was found that the torus was blown out over $\sim42$ inner orbit times. We compare this with Run a2\_e01, where the equatorial sublimation radius (which we define as radius where the initial dust fraction is $0.5$ of the large-distance value) is $R_s=0.01$ pc, which at a circular velocity of $\sim500$ km/s gives an orbital time of $120$ years, we find that the simulation runs for $\sim80$ inner orbit times, by which point the disk is still only being slowly destroyed. Even at high Eddington rates where $\dot{M}_w\sim0.005$~M$_\odot$yr$^{-1}$, the destruction time-scale for the disk $M/\dot{M}_w\sim13$ kyr, and this is solely considering the small $64$~M$_\odot$ central region of the disk that we are modeling. By contrast, a0\_e1 and a1\_e1 blow away the disk completely by the end of the simulation time. It is possible that introducing anisotropy to the radiation field in the model of \citet{2016ApJ...825...67C} may produce a longer-lasting more persistent torus, although removing flux from the equatorial plane may reduce radiation pressure support and cause the torus to collapse.

However, while the torus body moves outwards at late times in the isotropic model of \citet{2016ApJ...825...67C}, this is slower than the immediate blowout that we observe in our isotropic model. In that work, the simulations were equilibrated with a sub-keplerian rotation curve to maintain a longer lasting torus. We attempted to do this in our simulations, but found that because the outwards momentum from radiation pressure is deposited in the very thin skin region on the front of the disk, a sub-keplerian rotation curve did not produce a stable disk, but instead causes gas to pile up at the inner front, forming a dense self-gravitating ring. Our hypothesis is that two features in the model of \citet{2016ApJ...825...67C} smooth out the momentum deposition and allow stable sub-keplerian motion. Firstly, we use realistic values for the opacity of the dust-gas mix, while \citet{2016ApJ...825...67C} use a scaled-down UV opacity to spread out the optically thin region in order to resolve it. Secondly, radiation pressure from infrared photons emitted from the hot front can contribute to spreading radial momentum further out in the disk, and we have excluded the dust emission in this work. We intend to test this in our future paper, where we will include infrared radiation from the hot dust, but with realistic UV opacities.

We also find that supernovae do not appear to be efficient at launching material high above the disk, even with a very large supernovae rate. This is contrary to the approach of \citet{2016ApJ...828L..19W} who invoke a large supernova rate to push material high enough above the disk to produce the observed opening angles. Assuming an initial mass function that produces one supernova per $100$~M$_\odot$ of star formation, their supernova rate is the equivalent of a star formation density of $\sim6000$~M$_\odot$~yr$^{-1}$~kpc$^{-2}$, and the effective star formation rate of our run a2\_e01\_SN1000 is greater at $>10^5$ M$_\odot$~yr$^{-1}$~kpc$^{-2}$. In our simulations, the effect of supernovae is localized, and drives a small bubble of hot gas out of the disk, having little effect on surrounding gas. This raises the question of why supernovae are so effective in the simulations of \citet{2016ApJ...828L..19W}. We suggest two possible explanations. The first is that this may simply be a resolution issue. If the supernova bubble size is smaller than or comparable to the resolution, then the bubble/chimney process can not be resolved, and a supernova spreads out its energy smoothly over a larger volume/mass of gas instead, driving a high-density low-velocity outflow rather than a low-density high-velocity outflow. With a resolution of $0.125$ pc, this may be the case in \citet{2016ApJ...828L..19W}. Secondly, if the number of supernovae per orbit is large, supernovae do not act as discrete explosions expelling small packets of gas, but as a smooth force inflating the entire disk. The supernova rate of a2\_e01\_SN1000 is $0.001$~yr$^{-1}$ and the outer orbital period (at $0.47$ pc) is $\sim30000$~yr, giving a product of $30$ supernova per outer orbit. The supernova rate of \citet{2016ApJ...828L..19W} is $0.014$~yr$^{-1}$ and the orbital period at $\sim6$ pc is $\sim1$~Myr, giving a product of $14000$ supernovae per orbit, at only part-way out of the disk. That is, we should naturally expect the effects of supernovae to be smoother and stronger at larger length scales and longer time-scales, despite the somewhat lower star formation surface density.

The model images in the molecular emission lines and the dust maps show some qualitative agreement with observations. Specifically, the relatively thin disk of dust emission within the central parsec resembles the thin mid-IR emitting disks seen in mid-IR interferometry in Circinus and NGC 1068 \citep[e.g.][]{2014A&A...563A..82T,2014A&A...565A..71L}. It may be responsible for the $3-5$\,\micron emission bump seen in the broad-band SED of many type 1 AGN \citep[e.g.][]{1986ApJ...308...59E,2011A&A...536A..78K,2012MNRAS.420..526M,2013ApJ...771...87H}. VLTI/GRAVITY and VLTI/MATISSE will be able to spatially resolve this region in nearby AGN and test the presence of such a disk. The molecular maps of the inner parsec show a clear density and temperature stratification, with the HCN probing the coolest and highest-density regions ($\sim10^6$\,cm$^{-2}$), CO slightly lower densities ($\sim10^6$\,cm$^{-2}$) and H$_2$ higher temperatures than the previous lines. As all three lines will become observable on small scales around the AGN with ALMA's long baseline configuration (CO, HCN) and VLTI/GRAVITY or the ELT (H$_2(1-0)$), a comparison of the velocity dispersions of those line will probe the density stratifications in the disk and compare it to the model predictions.

\section{Conclusions}\label{section_conclusion}

By performing radiation hydrodynamics simulations of a dusty gas disk around an AGN source, and neglecting re-radiation of IR photons, our models produce biconical outflows that could explain the geometrical distribution of IR emission observed in IR interferometry, as well as the extinction of an AGN over a significant solid angle. The outflows have a fairly low covering fraction, because radiation pressure from IR re-radiation is neglected. We will consider the more computational complex problem of IR re-radiation in a future paper. Nevertheless, this first model has produced a number of interesting results:

\begin{itemize}

\item All simulations produced a two-phase structure consisting of a hot outflowing component and a cool rotating component. The hot/cold gas component corresponds to the hot/cold dust component, although the gas temperature is hotter than the dust temperature. This suggests that observational tracers of on species (gas or dust) in a certain temperature range (hot/cold) will also trace the other species (dust or gas).

\item We found that, at constant luminosity, the anisotropy of radiation field is the key factor in governing the dynamics of the outflow and disk is the flux in the plane of the disk. This emphasizes the importance of constraining the anisotropy of the radiation field of the central engine of the AGN - the outflow rate is sensitive to the anisotropy factor almost as much as it is sensitive to the Eddington factor.

\item The distribution of dust temperatures qualitatively matches IR interferometry, although the lack of IR-reradiation causes the polar emission to be flattened. Additionally, the dust temperatures in the outflow are proportional to the outflow velocity, while the gas temperatures tend to fall into discrete phases. This is caused by the tighter connection between the radiation field and the dust than between the radiation field and the gas. This connection between the dust temperature and velocity may be useful in interpreting unresolved IR observations.

\item Supernovae only have a modest effect in increasing the outflow rate and covering fraction of the wind, even with an unrealistically high star formation surface density of $>10^5$ M$_\odot$~yr$^{-1}$~kpc$^{-2}$.

\item The outflow properties are mostly insensitive to imposing a gas temperature floor, even a severe temperature floor of $3000$ K. This demonstrates that the outflows are dominated by radiative interactions and not by hydrodynamics, even if the higher temperature floor produces a slightly thicker disk.

\item The mock emission line and continuum images of the central parsec around the AGN predict the presence of a relatively thin disk with a strong vertical density and temperature stratification. This can be tested with ALMA, VLTI, and/or ELT observations and will provide crucial information on the degree of turbulence in the AGN environment.

\end{itemize}

\section*{Acknowledgements}This research is supported by European Research Council Starting Grant ERC-StG-677117 DUST-IN-THE-WIND.
\bibliographystyle{apj}
\bibliography{radpressure1}

\end{document}